# Cost-benefit Analysis And Comparisons For Different Offshore Wind Energy Transmission Systems

Jesus Silva-Rodriguez[1]
Department of Electrical and Computer Engineering
University of Houston
Houston, TX, USA
jasilvarodriguez@uh.edu

Jin Lu[1]
Department of Electrical and Computer Engineering
University of Houston
Houston, TX, USA
jlu27@cougarnet.uh.edu

Xingpeng Li
Department of Electrical and Computer Engineering
University of Houston
Houston, TX, USA
xli82@uh.edu

*Abstract*— **This paper investigates how to efficiently and economically deliver offshore energy generated by offshore wind turbines to onshore. Both power cables and hydrogen energy facilities are analyzed. Each method is examined with the associated proposed optimal sizing model. A cost-benefit analysis is conducted, and these methods are compared under different scenarios. Three long-distance energy transmission methods are proposed to transmit offshore energy onshore, considering the capital/operation costs for related energy conversion and transmission facilities. The first method that deploys power cables only is a benchmark method. The other two methods utilize a hydrogen supercenter (HSC) and transmit offshore energy to and onshore substation through hydrogen pipelines. For the second method, electrolyzers are placed at offshore wind farms connected to HSC with low-pressure hydrogen pipelines. For the third method, offshore wind power is distributed to the HSC and then converted into hydrogen using electrolyzers placed at the HSC. An offshore scenario including three wind farms located in the Gulf of Mexico and one onshore substation is used as the base test case. Numerical simulations are conducted over a planning period of 30 years using the proposed optimization models for each method separately. Simulation results show that the proposed hydrogen-based methods can effectively transmit offshore energy to the onshore substation. The distances between the wind farms can influence the selection and configuration of the offshore wind energy system. Moreover, the proposed study analyzes the cost and benefits for various systems with different energy carriers. It can also suggest which method could be the best candidate for offshore energy delivery system under which scenario and demonstrate a promising future for offshore wind power generation.**

## Parameters

| | |
|---|---|
| $N^{WF}$: | Total number of wind farm locations. |
| $C_t^{LMP}$: | Electricity locational marginal price. |
| $P_{lim,k}^L$: | Power limit of each transmission line. |
| $V^O$: | Voltage level at the receiving-end of each line. |
| $d_k^L$: | Distance of each transmission line. |
| $R^L$: | Resistance of each transmission line. |
| $\eta^{DC}$: | Efficiency of DC power converters. |
| $CP^E$: | Energy intensity of each electrolyzer. |
| $CP^{FC}$: | Energy intensity of each fuel cell. |
| $CP^C$: | Energy consumption ratio of each compressor. |
| $H_{lim,k}^P$: | Hydrogen flow limit of each pipeline. |
| $p^O$: | Output pressure of all pipelines. |
| $T'$: | Mean gas temperature. |
| $d_k^P$: | Distance of each pipeline. |
| $D^P$: | Diameter of each pipeline. |
| $HS_t$: | Storage level of hydrogen storage system. |
| $HS_{max}$: | Maximum hydrogen storage limit. |
| $HS_{min}$: | Minimum hydrogen storage limit. |
| $P_{rated,k}^E$: | Rated power of each electrolyzer. |
| $P_{rated}^{FC}$: | Rated power of each fuel cell. |
| $Y$: | Considered service life (years) of the offshore system. |

## Sets & Indices

| | |
|---|---|
| $T$: | Time period set. |
| $K$: | Offshore wind farms set. |
| $t$: | Time interval index. |
| $k$: | Wind farm location index. |

## Variables

| | |
|---|---|
| $VP_t^{Del}$: | Total power delivered to onshore substation. |
| $C^T$: | Total cost of offshore transmission network. |
| $P_{out,k,t}^L$: | Power output for each transmission line. |
| $N_k^L$: | Number of transmission lines from each wind farm. |
| $P_{in,k,t}^L$: | Power input for each transmission line. |
| $P_{in,k,t}^E$: | Power input of each electrolyzer. |
| $H_{out,k,t}^E$: | Hydrogen flow output of each electrolyzer. |
| $P_{out,t}^{FC}$: | Power output of each fuel cell. |
| $H_{in,t}^{FC}$: | Hydrogen flow input of each fuel cell. |
| $P_{k,t}^C$: | Power consumption of each compressor. |
| $H_{k,t}^C$: | Hydrogen flow produced by each compressor. |
| $H_{out,k,t}^P$: | Hydrogen output for each pipeline. |
| $N_k^P$: | Number of hydrogen pipelines from each wind farm. |
| $p_{k,t}^I$: | Input pressure of each pipeline. |
| $H_{in,t}^S$: | Hydrogen storage input flow. |
| $H_{out,t}^S$: | Hydrogen storage output flow. |
| $N_k^E$: | Total number of electrolyzers. |
| $N^{FC}$: | Total number of fuel cells. |

## I. INTRODUCTION

Wind power is currently one of the most efficient and fast-growing renewable sources, however, it is limited primarily by land space and competing site usage. This is a factor that can be resolved with offshore wind farms. Offshore wind has

---

experienced continuous growth in the last years, and this trend is expected to continue in the next future [1].

Despite the high efficiency and high wind power potential found offshore, one main challenge this technology presents is bringing the power into shore for distribution. Offshore wind farms can be connected to onshore grids via AC or DC power transmission, and the current common trend in research is that for relatively long distances High Voltage Direct Current (HVDC) is the option with lower losses as well as minimal impact on grid disturbances [2-3]. Moreover, a HVDC point-to-point connection via voltage source converters (VSC) offers the cheapest option in terms of capital cost, in addition to operational advantages such as voltage control for a controlled DC power flow [3].

Another way that power can be transmitted is in the form of hydrogen. This process is known as power-to-gas, in which electrical power is used to generate hydrogen gas to be used as an energy carrier. In this way, energy can be delivered using hydrogen gas as a fuel, which inherently can be used as a means of energy storage, representing a cost-effective solution that can compete with battery energy storage systems [4]. Also, studies show that the hydrogen integrated grids can reduce both operation costs and $CO_2$ emissions [5-6].

Based on current research trends and practices, two main paths for wind power transmission from offshore to onshore can be via HVDC transmission lines or hydrogen pipelines. Both methods of offshore power transmission have been studied and compared in literature [7-9], and the general consensus indicates that after certain distances from shore, hydrogen transmission becomes the most economical solution.

Both HVDC transmission and power-to-gas hydrogen transmission from wind farm to onshore were analyzed and compared in [7]. This paper concludes that hydrogen is less expensive than HVDC at distances larger than 740 km. However, the paper only considers power-to-gas on both cases, and there is no consideration of the advantage to use hydrogen as a means for energy storage. Moreover, some important factors such as leakage from hydrogen and drops in pressure throughout the pipelines were not considered, which can provide useful information when demonstrating which transmission technology provides the best economic benefits.

A more thorough investigation and comparison between long-distance energy transport via hydrogen pipelines and HVDC transmission is performed in [8]. Here the results indicate that overall, at long distances, hydrogen transmission yields a lower levelized cost of energy compared to HVDC offshore transmission. However, although this study does consider hydrogen energy storage, both cases involve the generation of hydrogen gas, either at the wind farm or onshore, thus the costs of both cases involve the capital costs of main components, making the comparison exclusively about the path of transmission, i.e., the power lines and the pipelines along with the components these require.

In [9], a more separate comparison between these two means of power transmission is performed, and the technologies involved with HVDC transmission and hydrogen transport via pipelines are considered separate. HVDC transmission does not involve any hydrogen generation, and hydrogen transmission only converts back to electrical power onshore and takes advantage of the energy storage aspect of hydrogen gas. However, [9] considers only transmission of power from one wind farm location to onshore, and the possibility of interconnecting multiple wind farm projects is not explored.

To further investigate under which conditions hydrogen transport via pipelines provides greater cost-benefits than HVDC power transmission as well as provide different configuration options for offshore transmission networks, this paper proposes an offshore hydrogen super grid topology that couples multiple offshore wind projects via a hydrogen super center (HSC) located offshore and transmits the combined power to an onshore substation.

Three different cases will be considered for comparison: a) HVDC case: direct point-to-point HVDC transmission from each wind farm location to an onshore substation; b) Hybrid case: point-to-point HVDC transmission from each wind farm location to a HSC offshore, then power transmission via high pressure hydrogen pipelines (HPHP) from the HSC to the onshore substation; and c) Hydrogen pipelines (HP) case: hydrogen generation at each wind farm location and transmission via low pressure hydrogen pipelines (LPHP) to the offshore HSC, then all hydrogen collectively transmitted via HPHP from the HSC to the onshore substation. Diagrams for these three configurations are given in Fig. 1.

The rest of the paper is organized as follows. Section II presents the optimization model that is used to carry out the cost-benefit comparison between the three different cases, Section III explains the model simulation of the three test cases and presents the results of this economic comparison between the three method of offshore power transmission. Finally, Section IV concludes the paper.

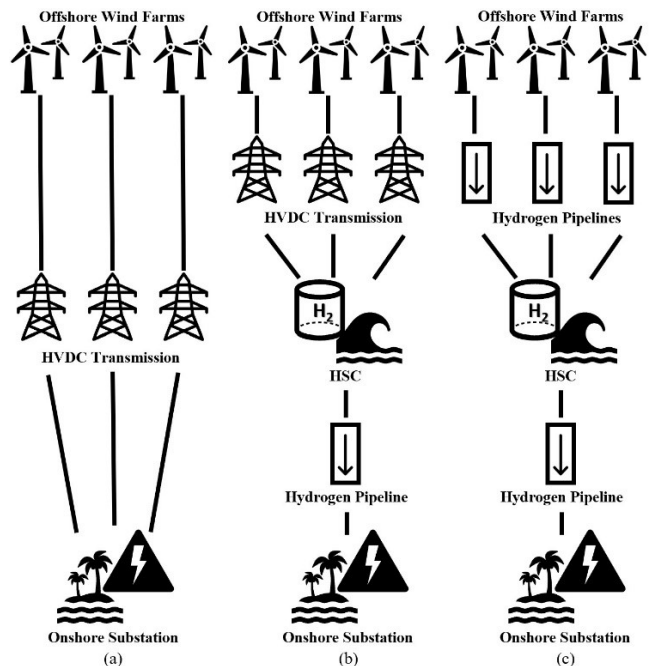

Fig. 1: Offshore power transmission comparison cases: (a) Benchmark point-to-point HVDC transmission, (b) Hybrid HVDC and hydrogen pipeline transmission, (c) Full offshore hydrogen pipeline transmission.

II. MODEL FORMULATION

The optimal planning for the three different cases being considered in this paper is carried out by an optimal sizing mixed integer quadratically constrained optimization problem.

This formulation will vary depending on the case as each will require different components. Still all three will derive from the same general objective and constraint expressions.

The objective of this model is to maximize the net cost-benefit of implementing the offshore wind power system following the respective configuration and selling the generated power at the onshore substation. Therefore, the objective function given by (1) consists of the product of the locational marginal price and the net effective power delivered at the onshore substation subtracted by the total capital and operation costs of all components involved in the configuration being considered.

$$\max \ 365 * Y * \sum_{t \in T} C_t^{LMP} P_t^{Del} - C^T \quad (1)$$

Both the HVDC and the hybrid cases involve power lines; therefore, these will include constraints that describe the limit of each power line as well as the number of lines needed running from each wind farm location (2). DC power flow is also included and is formulated based on the steady-state DC flow of power between two nodes (3). This power flow formulation intrinsically involves losses on the line, and the equation is formulated with converter power losses already factored in.

$$0 \le P_{out,k,t}^L \le P_{lim,k}^L N_k^L \ , \ \forall k = 1, \dots, N^{WF}, t \in T \quad (2)$$

$$\eta^{DC} P_{in,k,t}^L - P_{out,k,t}^L = 1000 \left(\frac{P_{out,k,t}^L}{V^O}\right)^2 \frac{d_k^L R^L}{N_k^L} \ , \ \forall k = 1, \dots, N^{WF}, t \in T \quad (3)$$

The Hybrid and HP cases must involve constraints that describe the energy efficiencies of electrolyzers (4) and fuel cells (5), power consumption of compressors (6), and number of pipelines and pipeline limits (7). The hydrogen flow in a pipeline (8) is formulated based on the equation derived in [10] and implemented in [11] for the design of hydrogen transmission networks. In addition, these two cases also involve hydrogen storage and constraints describing the storage level (9) and limits (10).

The decisions on the number of electrolyzers (11) and fuel cells (12) are obtained similar to the number of power lines and pipelines in parallel from (2) and (7), respectively, which depend on the limits and rated power of each component, increasing the number according to the needs determined by the optimal sizing model.

$$P_{in,k,t}^E = CP^E H_{out,k,t}^E \ , \ \forall k = 1, \dots, N^{WF}, t \in T \quad (4)$$
$$P_{out,t}^{FC} = CP^{FC} H_{in,t}^{FC}, \forall t \in T \quad (5)$$
$$P_{k,t}^C = CP^C H_{k,t}^C \ , \ \forall k = 1, \dots, N^{WF}, t \in T \quad (6)$$
$$0 \le H_{out,k,t}^P \le H_{lim,k}^P N_k^P \ , \ \forall k = 1, \dots, N^{WF}, t \in T \quad (7)$$
$$\left(p_{k,t}^I\right)^2 - (p^O)^2 = 5007.7 \frac{\lambda Z T'}{\rho^H} \left(\frac{H_{out,k,t}^P}{N_k^P}\right)^2 \frac{d_k^P}{(D^P)^5} \ , \quad (8)$$
$$\forall k = 1, \dots, N^{WF}, t \in T$$
$$HS_t = HS_{t-1} + \Delta t \cdot \left(H_{in,t}^S - H_{out,t}^S\right) \ , \ \forall t \in T \quad (9)$$
$$HS_{min} \le HS_t \le HS_{max} \quad (10)$$
$$P_{in,k,t}^E \le P_{rated,k}^E N_k^E \ , \ \forall k = 1, \dots, N^{WF}, t \in T \quad (11)$$
$$P_{out,t}^{FC} \le P_{rated}^{FC} N^{FC} \ , \ \forall k = 1, \dots, N^{WF}, t \in T \quad (12)$$

According to the above, the constraint formulations of the three cases are concluded in Table I. Moreover, all three cases are subject to the same objective function (1).

All the capacities and number of components are what will help determine the total costs based on the capital and operation costs reported in literature [2] [8] [9] [12-14]. As per the physical quantities associated with the pipeline characteristic and hydrogen gas featured in (8), the hydrogen compressibility factor $Z$ and density $\rho^H$ are based on the thermodynamic analyses and mathematical models for hydrogen flow from [15-16], and the coefficient of friction $\lambda$ is estimated based on the theory of the Colebrook-White equation for transitional flow presented in [15].

Table. I. Formulation of the Three Cases.

| Case Name | HVDC Case | Hybrid Case | HP case |
|---|---|---|---|
| Constraints | (2)-(3) | (2)-(12) | (4)-(12) |

### III. SIMULATIONS

For the test cases of these simulations, a substation located at the shore in Texas, U.S was selected. Three wind farms with 720MW of capacity each are located offshore in the gulf area. The onshore substation has a hydrogen energy storage system with $4 \times 10^5$ kg $H_2$ of capacity. The sample transmission systems of the three cases are illustrated in Fig. 2-4. The HVDC lines are in red, and the HPs are in green.

In the expected future scenario, the electrolyzer, fuel cell and hydrogen pipeline costs are halved. Under this assumption, the relationship between the total revenue (Billion $) and hydrogen-electric conversion round-trip efficiency are shown in Table II.

Table. II. Total Revenue (Billion $) vs. Hydrogen Round-trip Efficiency.

| Round-trip Efficiency | 37.7% (Current) | 40% | 50% | 60% | 70% | 80% | 90% |
|---|---|---|---|---|---|---|---|
| HVDC case | 17.2 | 17.2 | 17.2 | 17.2 | 17.2 | 17.2 | 17.2 |
| Hybrid case | 7.5 | 8.1 | 11.0 | 13.9 | 16.8 | 19.8 | 22.7 |
| HP case | 7.0 | 7.7 | 10.6 | 12.5 | 16.5 | 19.5 | 22.4 |

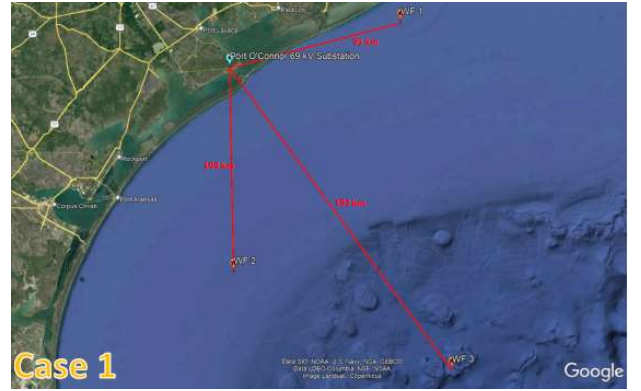

Fig. 2. Map of Transmission System of HVDC case.

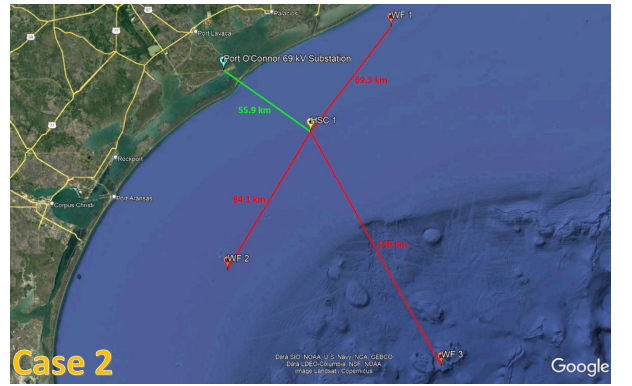

Fig. 3. Map of Transmission System of the Hybrid case.

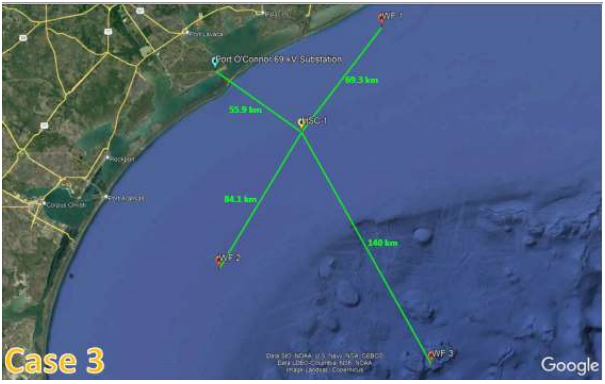

Fig. 4. Map of Transmission System of the HP case.

When the round-trip efficiency of the hydrogen-electric conversion is higher than 73%, the Hybrid case and HP case have a higher total revenue than the HVDC case. When the efficiency is 80%, the sizing results of HVDC lines, LPHPs, and HPHPs for different cases are shown in Table III.

Although the HVDC and Hybrid cases have the same number of HVDC lines, the total distance of HVDC lines in the Hybrid case is much lower than HVDC case. The reason is that the HVDC lines in the Hybrid case only deliver the electric power from the wind farms to the HSC. The distance between the HSC and wind farms is lower than the distance between the onshore substation and wind farms.

Table. III. The Sizing Results of Different Cases.

| Round Trip Efficiency | # of HVDC Lines | # of LPHP | # of HPHP | # of Electrolyzers | # of Fuel Cells |
|---|---|---|---|---|---|
| HVDC case | 3 | - | - | - | - |
| Hybrid case | 3 | - | 2 | 7,993 | 13,183 |
| HP case | - | 3 | 2 | 8,730 | 12,540 |

The hourly hydrogen-power exchanging conditions including the electrical energy consumed/generated by electrolyzers and fuel cells, and the energy stored in the hydrogen storage are plotted in Fig.5. and Fig. 6. for the Hybrid case and HP case separately. From the plots, we can observe that the Hybrid case and HP case have very similar hydrogen-power exchanging operations. The reason is that both models optimize the hourly energy exchange operations to maximize revenue.

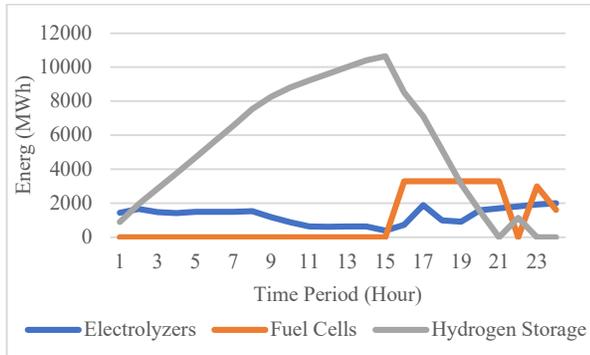

Fig. 5. Hydrogen-power Exchange of Hybrid Case.

In the second expecting scenario, we assume the round-trip efficiency can be improved to 70%. With this assumption, the relationship between the total revenue and electrolyzer/fuel cell costs is shown in Table IV. The results show that electrolyzer/fuel cell cost is an important factor affecting the total revenue.

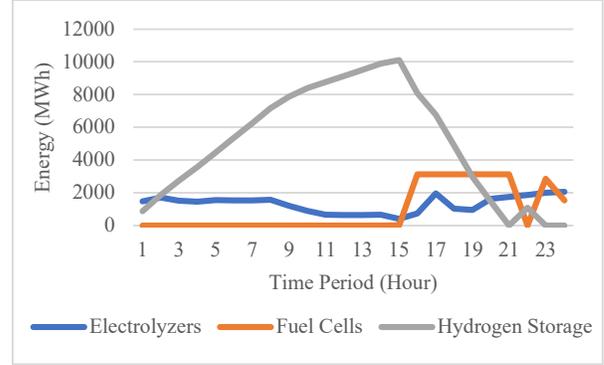

Fig. 6. Hydrogen-power Exchange of HP Case.

Table. IV. The Total Revenue (Billion $) vs. Electrolyzer/Fuel cell costs.

| Electrolyzer & Fuel Cell Cost Reduction | 0 | 20% | 40% | 60% | 80% |
|---|---|---|---|---|---|
| HVDC case | 17.2 | 17.2 | 17.2 | 17.2 | 17.2 |
| Hybrid case | 13.3 | 14.7 | 16.1 | 17.6 | 19.4 |
| HP case | 12.9 | 14.3 | 15.8 | 17.3 | 18.9 |

To study how the transmission distance may affect the performance of different cases, we scale the network in different proportions and run the simulations. The assumptions are that the fuel cell/electrolyzer cost and hydrogen line cost are halved. The round-trip efficiency is 70%. The revenue for different transmission distances is concluded in Table V, and the plot of revenue vs. network distance is shown in Fig. 7. We can observe that for longer distances, the pure hydrogen system has higher revenue. Both HVDC case and Hybrid case have obviously revenue reduction due to the existing of long distance HVDC lines.

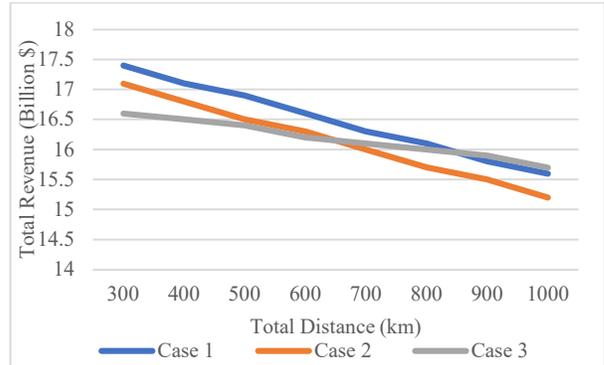

Fig. 7. Total Revenue vs. Total Transmission Distance.

Table. V. The Total Revenue (Billion $) vs. Transmission Distance.

| Distance from Wind Farms to Substation (km) | 300 | 400 | 500 | 600 | 700 | 800 | 900 | 1000 |
|---|---|---|---|---|---|---|---|---|
| HVDC case | 17.4 | 17.1 | 16.9 | 16.6 | 16.3 | 16.1 | 15.8 | 15.6 |
| Hybrid case | 17.1 | 16.8 | 16.5 | 16.3 | 16.0 | 15.7 | 15.5 | 15.2 |
| HP case | 16.6 | 16.5 | 16.4 | 16.2 | 16.1 | 16.0 | 15.9 | 15.7 |

We also study the influence of the wind farm capacity to the different cases. We select the future scenario with halved electrolyzer/fuel cell and hydrogen pipeline cost. The round-trip efficiency is 70%. The total revenue vs. wind farm capacity is shown in Table VI. We can observe that with wind farm capacity increases, both Hybrid case and HP case have much lower total revenue than HVDC case. The increased wind farm capacity leads to more wind energy generation. The Hybrid case and HP case require more hydrogen-power exchanging through electrolyzers and fuel cells. Due to the energy dissipated in the conversion, the Hybrid case and HP case will sell less energy to the bulk grid and have less total revenue. Hence, the limitation of the hydrogen integrated offshore transmission configuration is caused by the energy consumption in the hydrogen-electric conversion.

Table. VI. The Total Revenue (Billion $) vs. Wind Farm Capacity.

| Each Wind Farm Capacity (MW) | 360 | 720 | 1,080 | 1,440 | 1,800 |
|---|---|---|---|---|---|
| HVDC case | 7.7 | 17.2 | 24.9 | 34.4 | 43.8 |
| Hybrid case | 7.5 | 16.8 | 23.1 | 28.2 | 33.3 |
| HP case | 7.8 | 16.5 | 24.6 | 29.4 | 33.5 |

IV. CONCLUSIONS

This paper explored three different topologies of offshore wind power transmission to shore in order to compare and analyze the benefits of implementing power transmission in the form of hydrogen gas via pipelines. The three cases are a HVDC transmission, a mix of HVDC and hydrogen pipelines, and purely hydrogen pipeline transmission. The results of the analysis and comparisons demonstrate that, assuming a fixed reduction in cost of 50% in the future, an increase in power-gas-power round up efficiency up to 73% shows that both Hybrid and HP case will incur a better total cost than the HVDC case, with the HP case being slightly more economical. Moreover, with a fixed assumed increase up to 70% round-up efficiency the HVDC case is still economically better than the other two cases, showing that it is critical for these systems to not only achieve higher efficiencies, but also a cost reduction in electrolyzer and fuel cell components, in order to have these hydrogen transmission topologies perform better than offshore HVDC transmission. Moreover, the simulation results show that revenue improves for both the Hybrid case and the HP case compared to the HVDC case after around 1000 km, similar to what is seen in literature. This confirms that current tendencies place hydrogen power transmission topologies only more suitable than HVDC when wind power is being harvested deep into the ocean. However, this study has successfully determined that if current improvement trends of electrolyzer/fuel cell efficiencies and costs continue to improve, there will be cases in the future in which either a mix of hydrogen and HVDC or pure hydrogen transmission will be more economical at just around 100 km.